# Ambient pressure Dirac electron system in quasi-two-dimensional molecular conductor α-(BETS)$_2$I$_3$

Shunsuke Kitou[1,*,†], Takao Tsumuraya[2,‡,†], Hikaru Sawahata[3], Fumiyuki Ishii[4], Ko-ichi Hiraki[5], Toshikazu Nakamura[6], Naoyuki Katayama[1], and Hiroshi Sawa[1,#]

[1]*Department of Applied Physics, Nagoya University, Nagoya 464-8603, Japan*
[2]*Priority Organization for Innovation and Excellence, Kumamoto University, 2-39-1 Kurokami, Kumamoto 860-8555, Japan*
[3]*Graduate School of Natural Science and Technology, Kanazawa University, Kanazawa 920-1192, Japan*
[4]*Nanomaterials Research Institute, Kanazawa University, 920-1192 Kanazawa, Japan*
[5]*Department of Natural Sciences, Fukushima Medical University, Fukushima 960-1295, Japan*
[6]*Institute for Molecular Science, Myodaiji, Okazaki 444-8585, Japan*

We investigated the precise crystal structures and electronic states of a quasi-two-dimensional molecular conductor α-(BETS)$_2$I$_3$ at ambient pressure. The electronic resistivity of this molecular solid shows metal-to-insulator (MI) crossover behavior at $T_{MI}$ = 50 K. Our x-ray diffraction and $^{13}$C nuclear magnetic resonance experiments revealed that α-(BETS)$_2$I$_3$ maintains the inversion symmetry below $T_{MI}$. First-principles calculations found a pair of anisotropic Dirac cones at a general $k$-point, with the degenerate contact points at the Fermi level. The origin of the insulating state in this system is a small energy gap of ~2 meV opened by the spin–orbit interaction. The Z$_2$ topological invariants indicate that this system is a weak topological insulator. Our results suggest that α-(BETS)$_2$I$_3$ is a promising material for studying the bulk Dirac electron system in two dimensions.

## I. INTRODUCTION

A massless Dirac electron system, in which two linear band dispersions intersect at the Fermi level ($E_F$), is one of the central themes of modern condensed-matter physics [1-7]. When a system has such an emergent band structure, the electron behavior, such as electronic transport, follows the Dirac equation, and the charge carriers move at the speed of light in a material as if they had no mass. However, there are only a very few true massless Dirac electron systems, i.e., material systems in which the Dirac point is located at the $E_F$ and the band gap is zero. Such an electronic state is realized in a two-dimensional (2D) layer of graphene [1], in bismuth [8-10], and on the surface of topological insulators [11,12]. It has been suggested that unusual phenomena such as the quantum Hall effect [1], quantum spin Hall effect [2,4], and unscreened long-range Coulomb interaction [7] attributed to the Dirac cone band structure can be observed in these systems. In addition, applications to high-mobility electronic devices may exist [13,14].

Recently, massless Dirac electron systems have been shown to exist in some organic molecular solids [15-33], with Dirac cones formed by the bands of the same character of wavefunctions as frontier orbitals of consistent molecules at different sites. Such a massless Dirac electron system "in bulk" was first realized in a quasi-2D molecular conductor, α-(ET)$_2$I$_3$ [ET = BEDT-TTF = bis(ethylenedithio)tetrathiafulvalene] [Fig. 2(a)] [17-21], which, unlike graphene [1], has a pair of anisotropic Dirac cones [17-21]. However, the massless Dirac state in α-(ET)$_2$I$_3$ is realized only under high-pressure ($P$ > 1.2 GPa) [33]. At ambient pressure and $T_{MI}$ = 135 K, α-(ET)$_2$I$_3$ shows a metal–insulator (MI) transition, which causes a charge ordering (CO) associated with the lack of an inversion center, and the system turns to a nonmagnetic ferroelectric phase [34-42]. Further, the CO transition can be suppressed by applying pressure, and an anomalous electronic conducting phase, including a massless Dirac electron system, can be realized under high-pressure [16]. Although the quantum Hall effect [28], discrete Landau levels [29], and unscreened long-range Coulomb interactions [31,32] are observed under high-pressure in α-(ET)$_2$I$_3$, experimental determination of the detailed crystal structure and physical property measurements in the Dirac state are still limited.

To address the limitations mentioned above, we searched for a bulk Dirac electron system realized at ambient pressure. We found a promising candidate in the selenium-substituted analog of α-(ET)$_2$I$_3$, α-(BETS)$_2$I$_3$ [BETS = BETS-TSF = bis(ethylenedithio)tetraselenafulvalene] [Fig. 2(b)], where the central four S atoms in the ET molecule are replaced by Se atoms. The resistivity of α-(BETS)$_2$I$_3$ behaves like that of α-(ET)$_2$I$_3$, and the MI crossover temperature of α-(BETS)$_2$I$_3$ ($T_{MI}$ = 50 K) [43] is less than the CO transition temperature of α-(ET)$_2$I$_3$ [34]. As the temperature decreases from room temperature to low-temperature (LT), the magnetic susceptibility of α-(BETS)$_2$I$_3$ gradually decreases, and no anomaly exists at $T_{MI}$ [44]. These elec-tronic properties are different from those of α-(ET)$_2$I$_3$, and the MI crossover instead of the CO transition seems to occur as the temperature is decreased. The origin of the insulating state in α-(BETS)$_2$I$_3$ has not so far been understood.

In a previous theoretical study using the structure of α-(BETS)$_2$I$_3$ at room temperature and 0.7 GPa [24], a semimetallic band structure was obtained from first-principles density-functional-theory (DFT) method [20]. Tight-binding band structure calculations with extended Hückel parameters have failed to provide the zero-gap state (ZGS); they show different shapes of Fermi surface due to the over-tilting of Dirac cones [24,25]. Ambient-pressure structural and electronic properties, including atomic coordinates, have yet to be clarified; previous x-ray diffraction (XRD) studies at ambient-pressure provide only the lattice parameters and the space group at room temperature [24,43].

To verify the existence of the ZGS with a bulk Dirac electron system in α-(BETS)$_2$I$_3$ at ambient pressure, we investigate the crystal structures and electronic states by performing synchrotron XRD and $^{13}$C nuclear magnetic resonance (NMR) experiments. We find no clear phase transition in either experiment. We perform first-principles DFT calculations on this structure at LT. Our results strongly suggest the existence of a ZGS with bulk Dirac cone-type band dispersion in α-(BETS)$_2$I$_3$ at ambient pressure. The band gap of ~2 meV is opened by the spin–orbit coupling (SOC) effect. Finally, we discuss the difference in electronic structure between α-(BETS)$_2$I$_3$ and α-(ET)$_2$I$_3$.

## II. METHODS
### A. XRD experiments

XRD experiments were performed using a BL02B1 beamline at the synchrotron facility SPring-8 [45] in Japan. The dimensions of the α-(BETS)$_2$I$_3$ and α-(ET)$_2$I$_3$ crystals for the XRD experiments were $150 \times 150 \times 15\ \mu m^3$ and $140 \times 90 \times 20\ \mu m^3$, respectively. A helium-gas-blowing device was employed to cool the samples to 30 K. A 2D imaging-plate was used as the detector. The wavelength of the x-ray was 0.39054 Å, avoiding energy absorption at the $K$-edge of iodine (0.3738 Å). For the crystal structural analysis, we used original software for extracting the diffraction intensity [46]. SORTAV [47] and Jana2006 [48,49] were used for diffraction intensity averaging and crystal structural analysis, respectively.

### B. $^{13}$C NMR experiments

Single crystal $^{13}$C NMR measurements were performed in the same way as in an earlier study of α-(ET)$_2$I$_3$ [50]. The central double-bonded carbon atoms in BETS were selectively enriched with $^{13}$C isotope. An NMR spectrum was obtained by the fast Fourier transformation of the spin-echo signal induced by a π/2-π pulse sequence. The assignment of each peak to the molecular site was performed as follows. First, we measured the NMR spectrum in the **ab**-plane in the metallic state, and we found that the angular dependence of the peak positions was identical to that in α-(ET)$_2$I$_3$. This is reasonable because the molecular arrangements in the unit cell of α-(BETS)$_2$I$_3$ are similar to α-(ET)$_2$I$_3$. The peak assignments were quickly done in the **ab**-plane. Then, we tilted the field direction from the **a**-axis to the **c**-axis, keeping the peak assignments. The temperature dependence of the NMR spectrum was obtained in the field direction **B ∥ c** in which the chemical shift was reported to be sensitive to the fractional molecular charge in the case of α-(ET)$_2$I$_3$ [51].

### C. First-principles calculations

The present first-principles DFT calculations [52,53] are based on the exchange-correlation functional of generalized gradient approximation (GGA) proposed by Perdew, Burke, and Ernzerhof (PBE) [54]. For scalar-relativistic calculations, Kohn–Sham equations are self-consistently solved using an all-electron full-potential linearized augmented plane wave (FLAPW) method [55]. We also performed the calculations with a scheme based on plane waves and pseudopotentials generated by the projected augmented wave (PAW) formalism [56], which was implemented in Quantum Espresso (Q.E.) 6.3 [57,58]. The dimensions of the **k**-point meshes used were $6 \times 6 \times 2$ for the self-consistent loop, and $14 \times 14 \times 2$ and $16 \times 16 \times 2$ for the density of states without and with SOC, respectively. The results of both methods agreed well. Also, we performed nonmagnetic band structure calculations, including the SOC effect with full-relativistic pseudopotentials. Further, we calculate $Z_2$ topological invariants using OpenMX code [59–61]. The detailed computational conditions are summarized in Supplemental Material (SM) [62].

## III. RESULTS
### A. Crystal structure and NMR spectra

First, we investigated the crystal structure of α-(BETS)$_2$I$_3$ in the high-temperature (HT) phase at 80 K. To determine the bond length in the molecule (i.e., the amount of the molecular charge) with high accuracy, we performed a high-angle analysis; this is an effective method for the analysis of molecular crystals [63]. The structural analysis shows that α-(BETS)$_2$I$_3$ and α-(ET)$_2$I$_3$ have similar crystal structures in the HT phase [Figs. 1(a) and 1(b)], and the space group of both is $P\bar{1}$. There are four BETS molecules (A, A', B, and C) in a unit cell. Because there are inversion points at the centers of the mol-ecule A and A', these two are crystallographically equivalent. Our obtained lattice parameters of α-(BETS)$_2$I$_3$ generally agree with those reported in the previous structural studies at ambient pressure [43] and under a pressure of 0.7 GPa [24]. However, internal coordinates at ambient pressure have never been reported. Thus, here we fully determine structural parameters, including the lower temperature region. In SM [62], we show the results of the detailed analysis and compare the crystal structure we obtained at ambient pressure to that at 0.7 GPa reported in Ref. [24].

To investigate any potential changes in symmetry at $T_{MI}$, $^{13}$C NMR experiments were performed on α-(BETS)$_2$I$_3$. Figure 1(c) shows the NMR spectra for α-(BETS)$_2$I$_3$ at 100 K and 30 K obtained with the field direction $B \parallel c$. Maximum three doublets were observed even at 30 K. The splitting of the peak of the molecules A and A' due to the lack of an inversion center, which was observed at $T_{MI}$ for α-(ET)$_2$I$_3$ [51], is not observed in the LT phase of α-(BETS)$_2$I$_3$.

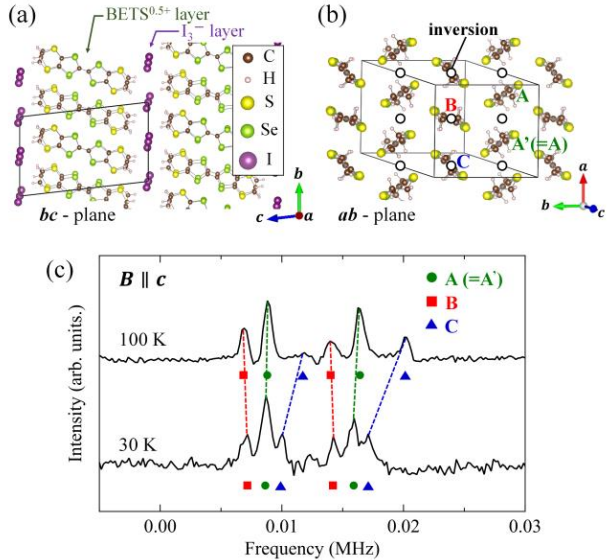

FIG. 1. Crystal structure of α-(BETS)$_2$I$_3$ in (a) $bc$-plane and (b) $ab$-plane. (c) $^{13}$C NMR spectra for α-(BETS)$_2$I$_3$ at 100 K and 30 K. An external field of 7 T was applied parallel to the $c$-axis. Zero frequency corresponds to the zero-Knight shift frequency.

In the LT phase of α-(BETS)$_2$I$_3$, additional superlattice reflections and/or splitting of the diffraction peaks were not confirmed from the XRD data. In addition, we did not find the lack of the inversion center even at 30 K from the structural analysis (Fig. S2 [62]), which is consistent with the result of $^{13}$C NMR measurement [Fig. 1(c)] and the past report of $^{77}$Se NMR measurement [44]. Therefore, we conclude that the space group is $P\bar{1}$ in the LT phase.

Next, we discuss the temperature dependence of the charge amount based on the bond length of the constituent BETS molecules in α-(BETS)$_2$I$_3$. As references, XRD experiments at SPring-8 and high-angle analysis were performed on α-(ET)$_2$I$_3$. Figure 2(c) shows the experimental evaluation of the charge amount $Q$ of α-(ET)$_2$I$_3$, which is calculated from the intra-molecular bond lengths corresponding to the C=C and C–S bonds [64]. The definition of $Q$ is given in the inset of Fig. 2(c). $Q$ varied largely due to the MI transition with the lack of an inversion center. In the LT phase, molecules A' and B are hole-rich, and molecule A and C are hole-poor, suggesting the existence of a horizontal-stripe-type CO state. This result is consistent with the results of several previous studies using infrared spectra [39], Raman spectra [40], NMR [41], XRD [42], and various theoretical calculations [20,36-38,65].

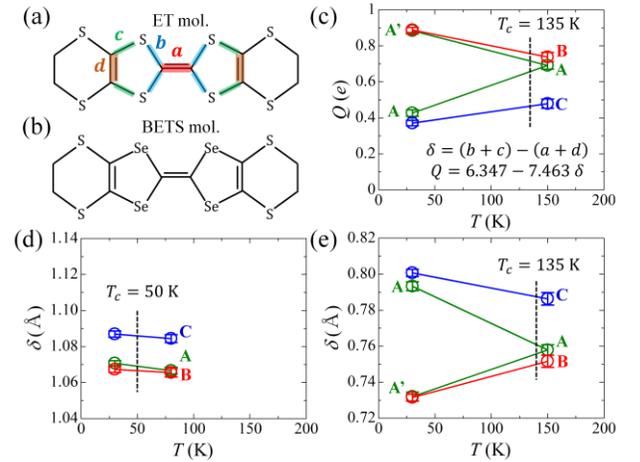

FIG. 2. (a),(b) Molecular structures of ET and BETS, respectively. (c) Temperature dependence of the charge amount $Q$ in ET [64] in α-(ET)$_2$I$_3$. (d),(e) Temperature dependence of the $\delta$ value in BETS and ET in α-(BETS)$_2$I$_3$ and α-(ET)$_2$I$_3$, respectively. The $\delta$ value [$\delta = (b + c) - (a + d)$] corresponds to the difference in length between the C=C and C–S bonds in the molecule.

Next, in Fig. 2(d), we show the temperature dependence of the charge amount on each BETS molecule in α-(BETS)$_2$I$_3$. Because the formula $Q$ for calculating the valence has not been reported for the BETS salt, a comparison is made using the $\delta$ value itself [inset of Fig. 2(c)]. For reference, Fig. 2(e) shows the temperature variation of $\delta$ for α-(ET)$_2$I$_3$. For α-(BETS)$_2$I$_3$, the amount of charge of BETS shows less change due to the MI crossover. The changes in bond length within BETS are less than 0.2% between 80 K and 30 K. Furthermore, the changes in the distance and angle between the BETS molecules are also insignificant (Figs. S4 and S5 [62]).

In previous work on α-(ET)$_2$I$_3$, changes of distances between donor molecules and terminal iodine atoms of I$_3$ before/after the phase transition associated with the CO were noted [20]. We therefore investigated the I–H distances in α-(ET)$_2$I$_3$ and α-(BETS)$_2$I$_3$. In α-(ET)$_2$I$_3$, apparent changes of the I–H distances are confirmed at $T_{MI}$, which is consistent with the previous report [20], whereas no changes are seen at $T_{MI}$ in α-(BETS)$_2$I$_3$ (Fig. S6 [62]). This result also shows the absence of CO in the LT phase of α-(BETS)$_2$I$_3$. These structural analysis results imply that symmetry and intra- and inter-molecular structures hardly change at the MI crossover of α-(BETS)$_2$I$_3$.

### B. Electron density distribution

In molecular solids, the valence and conduction bands, which control the physical properties, are made up of frontier orbitals of the constituent molecules [66]. Here, we focused on the valence electron density (VED) distribution to investigate the difference between α-(ET)$_2$I$_3$ and α-(BETS)$_2$I$_3$. An electron density (ED) analysis using a core differential Fourier synthesis (CDFS) method [63,67], which efficiently extracts only the valence electron contribution, was performed on these two compounds (see Ref. [67] for details about CDFS analysis).

We compared the VED distributions of molecule A in α-(ET)$_2$I$_3$ and α-(BETS)$_2$I$_3$, as shown in Fig. 3. Here, the valence electron configurations of the C, S, and Se atoms constituting the ET and BETS molecules are $2s^22p^2$, $3s^23p^4$, and $4s^24p^4$, respectively. The VED distributions of molecules A', B, and C were approximately identical to that of molecule A in real space (Figs. S8 and S9 [62]). In the HT phase, differentiating the contribution of the thermal vibrations from the VED is difficult because of the large temperature contribution. Indeed, the VED distribution of ET at 150 K is blurred [Fig. 3(a)]. However, relatively localized VED distributions are observed at 80 K [Fig. 3(b)] and 30 K [Figs. 3(c) and 3(d)]. It should be noted that there is a trade-off relationship between the resolution of XRD data and the statistical error of the weak reflection intensity. We confirmed the reliability of the VED distributions by changing the resolution (Fig. S10 [62]).

A change in the VED is observed between the CDFS analysis results at 80 K and 30 K [Figs. 3(b) and 3(d)]. The ED around Se sites is higher in the LT phase than that in the HT phase, although there is no significant change in the structural parameters going from above to below the MI crossover temperature. Comparing ET and BETS at 30 K, the ED near the C=C bonds in BETS is higher than that in ET. Furthermore, although the number of valence electrons is the same (6$e$ per atom), the VED around Se sites in BETS [Fig. 3(d)] is higher than that around S sites near the center in ET [Fig. 3(c)]. These results indicate that the VED is more localized in BETS than in ET. Considering the electronegativity, since the value of S atom is slightly larger than Se, our result is not well explained with the difference of electronegativity.

On the other hand, the difference in the ionic radius of the two elements may affect their VEDs. The $p$ orbitals of Se in α-(BETS)$_2$I$_3$ are higher in energy than those of S in α-(ET)$_2$I$_3$; the $p$ orbitals of Se (4$p$ state) are more delocalized (extended in space) than those of S (3$p$ state). As will be described later, this difference appears in bandwidths, as plotted in the total density of states (DOSs) (Fig. 6). In addition, the intermolecular transfer integrals of α-(BETS)$_2$I$_3$ [68] calculated by the tight-binding model certainly have larger values in the far range than those of α-(ET)$_2$I$_3$ [18]. However, because the molecular orbital-like valence charge densities formed by the mixing of several atomic orbitals due to SOC are very complicated, the final orbital state of BETS is not trivial. We think that this issue is an open question and a challenge for future researches.

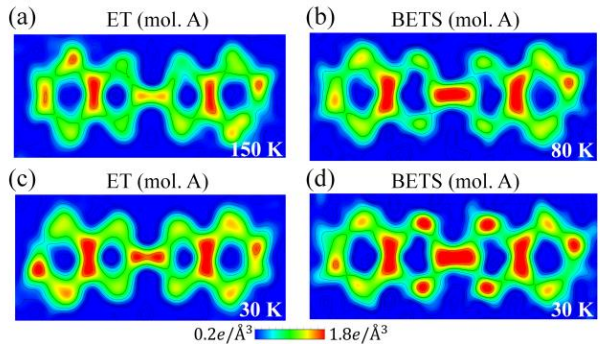

FIG. 3. Valence electron density distribution of molecule A in α-(ET)$_2$I$_3$ at (a) 150 K and (c) 30 K, and in α-(BETS)$_2$I$_3$ at (b) 80 K and (d) 30 K, obtained by the core differential Fourier synthesis analysis from the x-ray diffraction data in the limit $0\,\text{Å}^{-1} \leq \sin\theta/\lambda \leq 0.5\,\text{Å}^{-1}$.

### C. Band structure

Figures 4(a) and 4(b) show the calculated band structure and local density of states (LDOS) of α-(BETS)$_2$I$_3$ at 30 K without SOC. We find the Dirac cones are at general $\boldsymbol{k}$-points $(\pm 0.2958, \mp 0.3392, 0)$, not highly symmetric ones. No over-tilting of the Dirac cones is observed in Fig. 4(c). The LDOS is obtained as a summation of projected densities of state (PDOSs) on C $p$ and S $p$ states in the respective monomer units. The PDOSs are calculated within each muffin-tin sphere by FLAPW method. Using the LDOS from $-0.52$ eV up to the $E_F$, we compare the partition of holes on the individual molecule in the unit cell. The calculated values for molecules A and B are almost the same: 0.52 and 0.51, respectively. On the other hand, that for molecule C is smaller: 0.45. This tendency corresponds well to the $\delta$ value obtained from our structural data, as discussed in Sec. III.A [Fig. 2(d)]. These values are somewhat different from a previous DFT evaluation of hole distribution for the 0.7 GPa structure [20], where the donor A and C had similar charges, and the do-nor B was more positively charged.

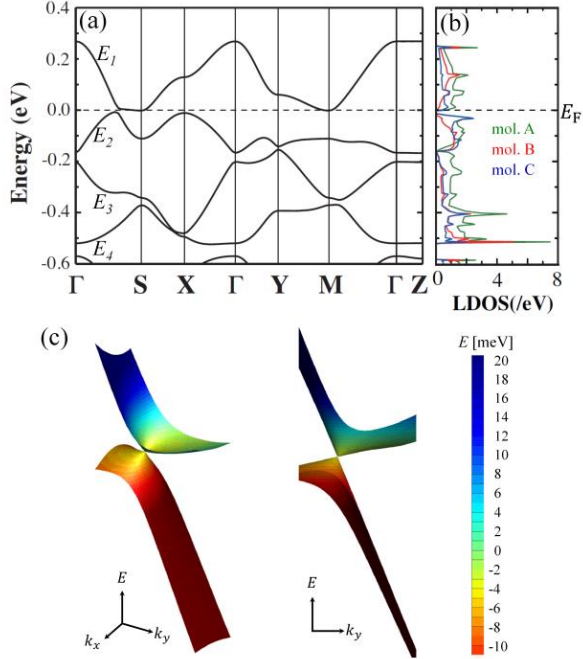

FIG. 4. (a) Band structure calculated from first-principles density-functional-theory and (b) local density of states (LDOS) of $\alpha$-(BETS)$_2$I$_3$ in the low-temperature phase (30 K) (without the spin–orbit coupling effect calculated with the FLAPW method). The dashed horizontal line shows the Fermi energy $E_F$. Green, red, and blue solid curves indicate the LDOS of molecules A, B, and C, respectively. (c) Band dispersion is seen from two directions close to the Dirac cone on the $\boldsymbol{k} = (k_x, k_y, 0)$ plane; a pair of Dirac points are located at $\boldsymbol{k} = (\pm 0.2958, \mp 0.3392, 0)$.

Near the $E_F$ in Fig. 4(b), the LDOSs of both A and C have a steep downward slope toward the $E_F$. On the other hand, the LDOS of B has a relatively gentle downward slope near the $E_F$. These trends in $\alpha$-(BETS)$_2$I$_3$ are similar to those in $\alpha$-(ET)$_2$I$_3$ at high-pressure [65]. This might be a consequence of the fact that the Se $p$ orbitals in $\alpha$-(BETS)$_2$I$_3$ are higher in energy than the S $p$ orbitals of $\alpha$-(ET)$_2$I$_3$; the $p$ orbitals of Se are more delocalized (extended in both energy and space) than those of S atoms. This leads to increasing bandwidth due to the Se substitutions to the TTF part; i.e., the upper band of $E_1$ in $\alpha$-(BETS)$_2$I$_3$ has a broader bandwidth (263 meV) than that of $\alpha$-(ET)$_2$I$_3$ (147 meV). The chemical substitution of Se atoms with S atoms thus plays a role similar to that of a physical pressure increase in $\alpha$-(ET)$_2$I$_3$. Interestingly, the insulating CO phase with $P1$ structure (no Dirac band dispersion) in $\alpha$-(ET)$_2$I$_3$ is suppressed by applying both uniaxial and hydrostatic pressures [16]. Therefore, we consider that controlling bandwidth is crucial for suppressing the CO transition and maintaining Dirac electron behavior even at LT.

In the no SOC limit, a massless Dirac electron system is realized when two linear bands intersect at the $E_F$ (where the Dirac points are located) [Fig. 5(a)]. However, with SOC, we observe a finite (indirect) energy gap of ~2 meV around the Dirac points, and the $E_F$ is located inside the energy gap, as plotted in Fig. 5(b). This insulator band dispersion is consistent with the increase in the electrical resistivity below $T_{MI} = 50$ K [43] and the decrease in the magnetic susceptibility with a decrease in temperature from room temperature [44]. Therefore, the insulating mechanism of $\alpha$-(BETS)$_2$I$_3$ is quite different from that of $\alpha$-(ET)$_2$I$_3$, showing the structural phase transition associated with CO [42]. We also note that the GGA functional slightly underestimates the band gap. The actual size of band gap is expected to be 4~5 meV, and the gap size also agrees well with the MI crossover temperature of ~50 K. However, because the energy gap of this system is quite small, unique physical properties, such as massless Dirac electron system behavior, are expected.

Next, we discuss the correspondence of the electronic structures of $\alpha$-(BETS)$_2$I$_3$ before and after the MI crossover at $T_{MI} = 50$ K. Figure 6(a) compares the magnified DOSs between 30 K and 80 K. Within an energy range from $-0.02$ eV to $0.02$ eV, including the band gap, the DOSs are almost the same. On the other hand, the DOSs outside this energy region are different, and those for the 80 K structures have slightly more expansive valleys. This difference originates with small changes in structural properties, e.g. inter-molecular distances (Fig. S4 [62]). In the band structure at 80 K, the band gap induced by SOC also exists. However, the thermal energy of 50 K is about 4.3 meV; the chemical potential can move over the energy gap and cut the finite DOSs. Therefore, we suggest that the electronic structure difference between 30 K and 80 K may contribute to the physical properties: the temperature effect on chemical potential explains the electronic conducting phase above $T_{MI}$ and the MI crossover behavior. A similar argument has been made previously about the DFT band structure of $\alpha$-(ET)$_2$I$_3$ calculated from an experimental structure measured at room temperature [18].

Based on the present results for $\alpha$-(BETS)$_2$I$_3$, we comment on the previous DFT studies of the crystal structure at a pressure of 0.7 GPa [24]. The energy band of $E_1$ close to the M(S) point is lower than the $E_F$. The lower band of $E_2$ close to the Y point is higher than the $E_F$, resulting in a semimetallic state. We have verified the above result for 0.7 GPa (with SOC), where the calculated DOSs shown in Fig. 6(c) indicate an entirely metallic state, although the overall band structure is consistent with ambient pressure results. In other words, the expected ground state at 0.7 GPa is different from that at ambient pressure.

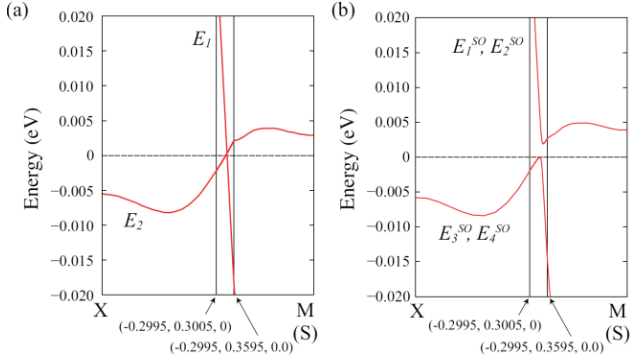

FIG. 5. Band dispersion of $\alpha$-(BETS)$_2$I$_3$ along the X (–0.5, 0, 0), (–0.2995, $y$, 0), and M(=S) (–0.5, 0.5, 0) lines (a) without and (b) with spin–orbit coupling near the Fermi energy $E_F$, calculated using Quantum Espresso code. The zero energies in (a) and (b) are set to be at the chemical potential and the top of the valence bands, respectively.

Next, we discuss the difference of electronic states from those for $\alpha$-(ET)$_2$I$_3$ above the CO transition temperature ($T_{CO}$). Above $T_{CO}$, $\alpha$-(ET)$_2$I$_3$ and $\alpha$-(BETS)$_2$I$_3$ crystals are isostructural. The ET salt has several common DOS features with the BETS salt, since $\alpha$-(ET)$_2$I$_3$ also has a Dirac cone-type band dispersion near the $E_F$ [18]. Figures 6(b) and 6(d) show the total DOS of $\alpha$-(BETS)$_2$I$_3$ at 30 K and $\alpha$-(ET)$_2$I$_3$ at 150 K, respectively. In both of these salts, associated with a Van Hove singularity close to the $E_F$, the DOSs show an asymmetric valley, centering on the zero energy. We find that the width of the valley in $\alpha$-(BETS)$_2$I$_3$ is much narrower than that in $\alpha$-(ET)$_2$I$_3$, although the width of each band is generally larger.

As discussed above, the bands forming the Dirac cones in $\alpha$-(BETS)$_2$I$_3$ are more flattened than those in $\alpha$-(ET)$_2$I$_3$. Therefore, the effective electron velocity of $\alpha$-(BETS)$_2$I$_3$ is expected to be smaller than that of $\alpha$-(ET)$_2$I$_3$. As shown in Fig. 6(b), the nearest peak above the chemical potential, located at +0.006 eV, is lower than that of +0.017 eV in $\alpha$-(ET)$_2$I$_3$ [Fig. 6(d)]. On the other hand, the nearest peak below $E = 0$ is located at –0.0075 eV in $\alpha$-(BETS)$_2$I$_3$, and is shallower than that in $\alpha$-(ET)$_2$I$_3$. The linear energy dependence of the DOSs attributed to the 2D Dirac cone is also much smaller than that in $\alpha$-(ET)$_2$I$_3$.

The delocalized character of Se $p$ orbitals causes such narrow energy windows close to the Dirac cones. In fact, the energy difference between eigenvalues is generally smaller than in $\alpha$-(ET)$_2$I$_3$. For instance, the energy difference between the highest occupied molecular orbital (HOMO) and the lowest unoccupied molecular orbital (LUMO) levels of an isolated BETS monomer is found to be 1.34 eV within the GGA-PBE functional; this is smaller than that for the isolated ET molecule (1.57 eV). Thus, the hybridization of wavefunctions with surrounding BETS molecules becomes more significant, making the number of relevant transfer energies much more extensive

[68]. The delocalized nature of Se $p$ orbitals also reduces the on-site Coulomb interaction $U$. The $U$ values calculated in the constrained random phase approximation are ~1.38 eV [69], and are generally smaller than those for $\alpha$-(ET)$_2$I$_3$ [70]. Recent NMR measurements of $1/T_1$ also confirmed the reduction of effective Coulomb interactions by Se substitution [71]. Therefore, we consider the delocalization of the molecular orbitals to reduce an excitonic instability (as discussed for $\alpha$-(ET)$_2$I$_3$ in Ref. [32]), and perhaps also to suppress the appearance of CO.

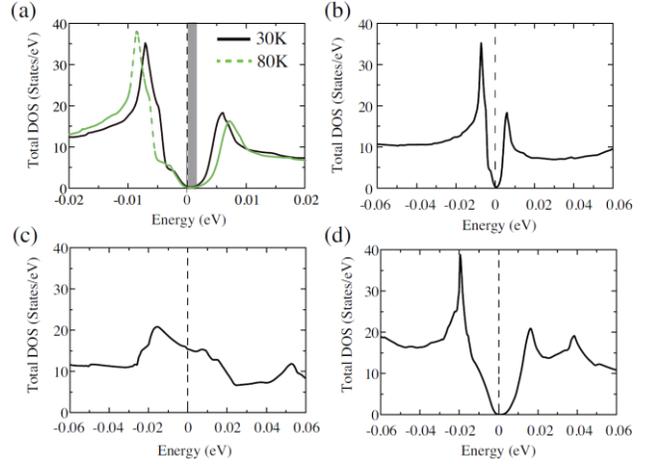

FIG. 6. (a) Total density of states (DOSs) close to the Dirac cones in $\alpha$-(BETS)$_2$I$_3$ at ambient pressure, when spin–orbit coupling is included. The solid (black) and dashed (green) curves show the DOSs at 30 K and 80 K, respectively. The grey shaded region lying above the energy zero (chemical potential) represents the band gap (~2 meV) in the 30 K structure. (b) DOS for the ambient pressure structure at 30 K (including the same data as the solid curve in (a) but plotted on a different scale). (c) DOS for the experimental structure under a pressure of 0.7 GPa [24]. (d) DOS of $\alpha$-(ET)$_2$I$_3$ for 150 K (above the charge-ordering transition temperature). The zero energies in (a), (b), and (d) are set to the tops of the valence bands; the zero energy in (c) is the Fermi energy $E_F$.

### D. $Z_2$ topological invariant

Barring a few reports [72-75], the SOC has been mostly ignored in theoretical studies for molecular solids because most of them are composed of light elements. However, we observe that the SOC critically changes the low-energy band structure from a massless Dirac electron system to a small band gap insulator. Therefore, to clarify whether the insulator state is a topological insulator or not, we have calculated $Z_2$ topological invariants from first-principles for bulk $\alpha$-(BETS)$_2$I$_3$ at 30 K using the parity eigenvalues at the time-reversal invariant momenta [76] and the parity method [11] (implemented in OpenMX code [77]). We have confirmed that the calculated topological invariants ($v$; $v_1 v_2 v_3$) are (0; 0 0 1), indicating a weak topological insulator. This suggests that an exotic massless Dirac band dispersion appears at the surface state along the *xz* direction.

To investigate whether the non-centrosymmetric CO phase in $\alpha$-(ET)$_2$I$_3$ is a topological insulating state, we have calculated the $Z_2$ topological invariant [76] for the experimental structure of $\alpha$-(ET)$_2$I$_3$ at 30 K using the Fukui-Hatsugai method [78] (implemented in OpenMX code [79]). The structural transition associated with CO is not always accompanied by the transition from a topological state to a trivial insulating state or vice versa. In this case, the calculated topological invariant for the CO phase is $Z_2 = 0$, implying a trivial insulator phase. On the other hand, the topological invariant for the HT phase with a centrosymmetric structure is $Z_2 = 1$, which indicates a topological semimetal phase [80]. These results suggest that the $Z_2$ topological phase transition occurs in $\alpha$-(ET)$_2$I$_3$ and is associated with the CO phase transition. We emphasize that the LT CO state of $\alpha$-(ET)$_2$I$_3$ is a trivial insulator in the ground state. By contrast, the band structure above $T_{MI} = 80$ K of $\alpha$-(BETS)$_2$I$_3$ remains that of a topological insulating state with (0; 0 0 1). Thus, no topological phase transition occurs in $\alpha$-(BETS)$_2$I$_3$, and the ground state is a weak topological insulator. Therefore, topological physical properties, such as the quantum spin Hall effect [4], can be observed in $\alpha$-(BETS)$_2$I$_3$, but not in the CO phase in $\alpha$-(ET)$_2$I$_3$. The calculated $Z_2$ topological invariants and the detailed computational conditions are summarized in SM [62].

## IV. SUMMARY

We studied the precise crystal and electronic structures of the quasi-2D molecular conductor $\alpha$-(BETS)$_2$I$_3$ at ambient pressure. Our XRD and $^{13}$C NMR measurements revealed that the crystal structure, unlike that of $\alpha$-(ET)$_2$I$_3$, is centrosymmetric even at 30 K, and the energy bands are Kramer's degenerate. To investigate the origin of the insulating state observed in electronic resistivity measurements, we performed first-principles calculations based on the crystal structure measured above and below the MI crossover temperature of ~50 K. At 30 K, we found linear crossing band dispersions close to $E_F$; we suggest a massless Dirac electron system with a ZGS is realized at ambient pressure. In contrast to the Dirac cone-type band structure in $\alpha$-(ET)$_2$I$_3$ above the CO transition temperature, the bands close to the Dirac point are comparatively flattened in $\alpha$-(BETS)$_2$I$_3$, and the band structure close to the Dirac cone is in a narrow energy window. Thus, the effective electron velocity of $\alpha$-(BETS)$_2$I$_3$ is expected to be smaller than that of $\alpha$-(ET)$_2$I$_3$, although overall bandwidth and transfer energies are generally larger.

Moreover, the degeneracies are removed by the SOC, resulting in an energy gap of ~2 meV near the Dirac points that corresponds well with the MI crossover temperature. Thus, the spin–orbit interaction converts $\alpha$-(BETS)$_2$I$_3$ from a zero-gap massless Dirac electron system to a weak topological insulator. This phenomenon is similar to that occurring in graphene, which has a smaller SOC-related energy gap ($\sim 0.8 \times 10^{-3}$ meV) than $\alpha$-(BETS)$_2$I$_3$ [81]. Thus, the quantum spin Hall effect is expected in $\alpha$-(BETS)$_2$I$_3$, as in graphene [2] and surface states of topological insulators. Our results have the potential to contribute significantly to the study of the Dirac electron system. In the near future, experimental results based on our expectations will be reported.

## ACKNOWLEDGMENTS

We thank N. Tajima, D. Ohki, A. Kobayashi, K. Yoshimi, Y. Suzumura, T. Naito, S. Fujiyama, R. Kato, and T. Takahashi for fruitful discussions. Crystal structure and ED distribution figures were visualized using VESTA [82]. CCDC 2008980–2008983 contain the supplementary crystallographic data for this paper, which are $\alpha$-(ET)$_2$I$_3$ at 150 K and 30 K, and $\alpha$-(BETS)$_2$I$_3$ at 80 K and 30 K. The data is provided free of charge by The Cambridge Crystallographic Data Centre [83]. This work was supported by a Grant-in-Aid for Scientific Research (Grants No. JP19J11697, and JP19K21860), JST CREST Grant No. JPMJCR18I2, and JSPS Research Fellow from JSPS. T.T. is partially supported by MEXT Japan, Leading Initiative for Excellent Young Researchers (LEADER). The computations were mainly carried out using the computer facilities of ITO at the Research Institute for Information Technology, Kyushu University, MASAMUNE at the Institute for Materials Research, Tohoku University, and the facilities of the Supercomputer Center, the Institute for Solid State Physics, the University of Tokyo. The synchrotron radiation experiments were performed at SPring-8 with the approval of the Japan Synchrotron Radiation Research Institute (JASRI) (Proposal No. 2017B1733).

* kitou.shunsuke@h.mbox.nagoya-u.ac.jp
‡ tsumu@kumamoto-u.ac.jp
# hiroshi.sawa@cc.nagoya-u.ac.jp
† These authors contributed equally to this work.

# Supplemental Material for

# Ambient pressure Dirac electron system in quasi-two-dimensional molecular conductor $\alpha$-(BETS)$_2$I$_3$

Shunsuke Kitou[1,*,†], Takao Tsumuraya[2,‡,†], Hikaru Sawahata[3], Fumiyuki Ishii[4], Ko-ichi Hiraki[5], Toshikazu Nakamura[6], Naoyuki Katayama[1], and Hiroshi Sawa[1,#]

*Department of Applied Physics, Nagoya University, Nagoya 464-8603, Japan*
[2]*Priority Organization for Innovation and Excellence, Kumamoto University, 2-39-1 Kurokami, Kumamoto 860-8555, Japan*
[3]*Graduate School of Natural Science and Technology, Kanazawa University, Kanazawa 920-1192, Japan*
[4]*Nanomaterials Research Institute, Kanazawa University, 920-1192 Kanazawa, Japan*
[5]*Department of Natural Science, Fukushima Medical University, Fukushima 960-1295, Japan*
[6]*Institute for Molecular Science, Myodaiji, Okazaki 444-8585, Japan*

[*]e-mail: kitou.shunsuke@h.mbox.nagoya-u.ac.jp
[‡]e-mail: tsumu@kumamoto-u.ac.jp
[#]e-mail: hiroshi.sawa@cc.nagoya-u.ac.jp
[†]These authors contributed equally to this work.

## Table of Contents

1. Crystal structural analysis
2. Electron density analysis
3. Computational details for electronic structures
4. Computational details for $Z_2$ topological invariants

# 1. Crystal structural analysis

The results of the crystal structural analysis of $\alpha$-(ET)$_2$I$_3$ and $\alpha$-(BETS)$_2$I$_3$ are shown in Fig. S1 and Table S1–S4. CCDC 2008980–2008983 contain the supplementary crystallographic data, which are $\alpha$-(ET)$_2$I$_3$ at 150 K and 30 K, and $\alpha$-(BETS)$_2$I$_3$ at 80 K and 30 K, for this paper. The data is provided free of charge by The Cambridge Crystallographic Data Centre.

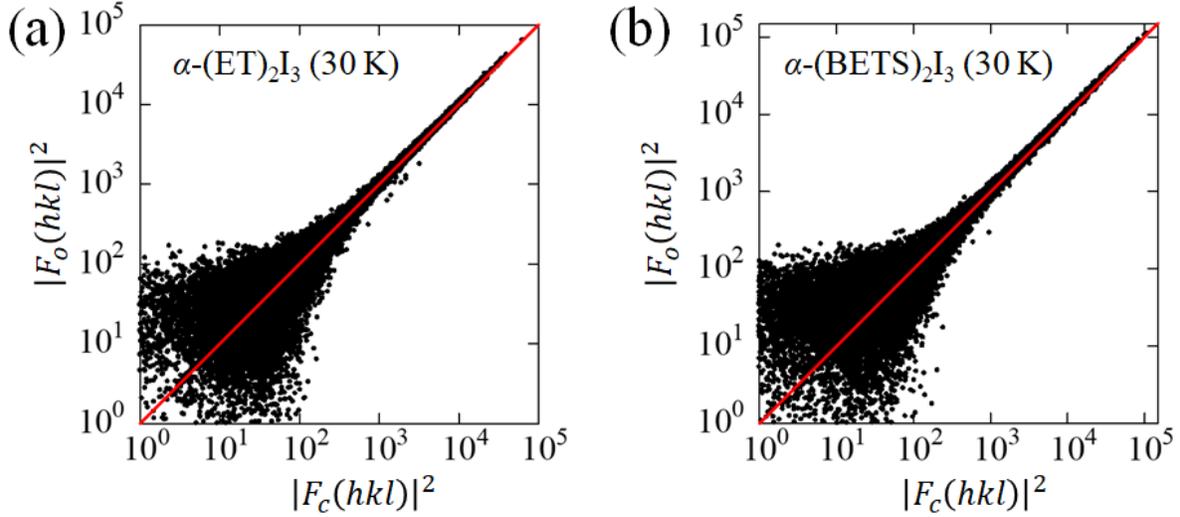

FIG. S1. Results of the crystal structural analysis (high-angle analysis) of (a) $\alpha$-(ET)$_2$I$_3$ at 30 K and (b) $\alpha$-(BETS)$_2$I$_3$ at 30 K using only high-angle reflections ($0.5$ Å$^{-1}$ $\leq \sin\theta/\lambda \leq 1.42$ Å$^{-1}$). $|F_c(hkl)|^2 - |F_o(hkl)|^2$ plots are shown as a double logarithmic. Here, $F_o(hkl)$ is the experimentally observed crystal structure factor, and $F_c(hkl)$ is the calculated crystal structure factor. Red line corresponds to $|F_c(hkl)|^2 = |F_o(hkl)|^2$. The number of reflections is (a) 90754 and (b) 65572, respectively.

Table S1. Summary of crystallographic data of $\alpha$-(ET)$_2$I$_3$ at the high-temperature phase.

| Chemical Formula | C20 H16 I3 S16 |
|---|---|
| Temperature (K) | 150 |
| CCDC deposit # | 2008980 |
| Wavelength (Å) | 0.39054 |
| Crystal dimension ($\mu$m$^3$) | 140 × 90 × 20 |
| Space group | $P\bar{1}$ |
| $a$ (Å) | 9.0983(18) |
| $b$ (Å) | 10.724(2) |
| $c$ (Å) | 17.388(4) |
| $\alpha$ (°) | 96.654(7) |
| $\beta$ (°) | 97.961(7) |
| $\gamma$ (°) | 90.864(8) |
| $V$ (Å$^3$) | 1668.0(6) |
| $Z$ | 2 |
| $F$ (000) | 1102 |
| $(\sin\theta/\lambda)_{Max}$ (Å$^{-1}$) | 1.42 |
| $N_{Total,obs}$ | 193468 |
| $N_{Unique,obs}$ | 63054 |
| Average redundancy | 3.1 |
| Completeness | 0.775 |
| High-angle analysis [0.5 Å$^{-1}$ ≤ sin $\theta$ /$\lambda$ ≤ 1.42 Å$^{-1}$] ($N_{Parameters}$ = 229) | |
| $R_1$ [# of reflections] | 0.1485 [44887] |
| $R_1$ ($I$ > 1.5$\sigma$) [# of reflections] | 0.0604 [22589] |
| GOF [# of reflections] | 1.04 [44887] |
| Normal analysis [0 Å$^{-1}$ ≤ sin $\theta$ /$\lambda$ ≤ 1.42 Å$^{-1}$] ($N_{Parameters}$ = 0) | |
| $R_1$ [# of reflections] | 0.1131 [48039] |
| $R_1$ ($I$ > 3$\sigma$) [# of reflections] | 0.0328 [19805] |
| GOF [# of reflections] | 1.07 [48039] |

Table S2. Summary of crystallographic data of $\alpha$-(ET)$_2$I$_3$ at the low-temperature phase.

| Chemical Formula | C20 H16 I3 S16 |
|---|---|
| Temperature (K) | 30 |
| CCDC deposit # | 2008981 |
| Wavelength (Å) | 0.39054 |
| Crystal dimension ($\mu$m$^3$) | 140 × 90 × 20 |
| Space group | $P$1 |
| $a$ (Å) | 9.0352(2) |
| $b$ (Å) | 10.6734(2) |
| $c$ (Å) | 17.3547(12) |
| $\alpha$ (°) | 96.541(7) |
| $\beta$ (°) | 97.752(7) |
| $\gamma$ (°) | 91.216(6) |
| $V$ (Å$^3$) | 1646.40(13) |
| $Z$ | 2 |
| $F$ (000) | 1102 |
| $(\sin\theta/\lambda)_{Max}$ (Å$^{-1}$) | 1.42 |
| $N_{Total,obs}$ | 327060 |
| $N_{Unique,obs}$ | 130278 |
| Average redundancy | 2.5 |
| Completeness | 0.810 |
| High-angle analysis [0.5 Å$^{-1}$ ≤ sin $\theta$ /$\lambda$ ≤ 1.42 Å$^{-1}$] ($N_{Parameters}$ = 347) | |
| $R_1$ [# of reflections] | 0.0522 [90754] |
| $R_1$ ($I$ > 1.5$\sigma$) [# of reflections] | 0.0353 [74900] |
| GOF [# of reflections] | 0.88 [90754] |
| Normal analysis [0 Å$^{-1}$ ≤ sin $\theta$ /$\lambda$ ≤ 1.42 Å$^{-1}$] ($N_{Parameters}$ = 0) | |
| $R_1$ [# of reflections] | 0.0465 [97167] |
| $R_1$ ($I$ > 3$\sigma$) [# of reflections] | 0.0268 [70334] |
| GOF [# of reflections] | 0.91 [97167] |

Table S3. Summary of crystallographic data of $\alpha$-(BETS)$_2$I$_3$ at the high-temperature phase.

| Chemical Formula | C20 H16 I3 S8 Se8 |
|---|---|
| Temperature (K) | 80 |
| CCDC deposit # | 2008982 |
| Wavelength (Å) | 0.39054 |
| Crystal dimension ($\mu$m$^3$) | 150 × 150 × 15 |
| Space group | $P\bar{1}$ |
| $a$ (Å) | 9.0996(2) |
| $b$ (Å) | 10.7301(2) |
| $c$ (Å) | 17.7313(12) |
| $\alpha$ (°) | 96.356(7) |
| $\beta$ (°) | 97.739(7) |
| $\gamma$ (°) | 90.777(6) |
| $V$ (Å$^3$) | 1704.25(13) |
| $Z$ | 2 |
| $F$ (000) | 1390 |
| $(\sin\theta/\lambda)_{Max}$ (Å$^{-1}$) | 1.42 |
| $N_{Total,obs}$ | 205480 |
| $N_{Unique,obs}$ | 73729 |
| Average redundancy | 2.8 |
| Completeness | 0.886 |
| High-angle analysis [0.5 Å$^{-1}$ ≤ sin $\theta$ /$\lambda$ ≤ 1.42 Å$^{-1}$] ($N_{Parameters}$ = 229) | |
| $R_1$ [# of reflections] | 0.0751 [54921] |
| $R_1$ ($I > 1.5\sigma$) [# of reflections] | 0.0409 [38378] |
| GOF [# of reflections] | 0.94 [54921] |
| Normal analysis [0 Å$^{-1}$ ≤ sin $\theta$ /$\lambda$ ≤ 1.42 Å$^{-1}$] ($N_{Parameters}$ = 0) | |
| $R_1$ [# of reflections] | 0.0646 [58389] |
| $R_1$ ($I > 3\sigma$) [# of reflections] | 0.0274 [34639] |
| GOF [# of reflections] | 0.95 [58389] |

Table S4. Summary of crystallographic data of $\alpha$-(BETS)$_2$I$_3$ at the low-temperature phase.

| Chemical Formula | C20 H16 I3 S8 Se8 |
|---|---|
| Temperature (K) | 30 |
| CCDC deposit # | 2008983 |
| Wavelength (Å) | 0.39054 |
| Crystal dimension ($\mu$m$^3$) | 150 × 150 × 15 |
| Space group | $P\bar{1}$ |
| $a$ (Å) | 9.0922(2) |
| $b$ (Å) | 10.7221(2) |
| $c$ (Å) | 17.7377(12) |
| $\alpha$ (°) | 96.310(7) |
| $\beta$ (°) | 97.706(7) |
| $\gamma$ (°) | 90.794(6) |
| $V$ (Å$^3$) | 1702.49(13) |
| $Z$ | 2 |
| $F$ (000) | 1390 |
| $(\sin\theta/\lambda)_{Max}$ (Å$^{-1}$) | 1.42 |
| $N_{Total,obs}$ | 353053 |
| $N_{Unique,obs}$ | 77235 |
| Average redundancy | 4.6 |
| Completeness | 0.929 |
| High-angle analysis [0.5 Å$^{-1}$ ≤ sin $\theta$ /$\lambda$ ≤ 1.42 Å$^{-1}$] ($N_{Parameters}$ = 229) | |
| $R_1$ [# of reflections] | 0.0554 [65572] |
| $R_1$ ($I$ > 1.5$\sigma$) [# of reflections] | 0.0359 [51466] |
| GOF [# of reflections] | 1.03 [65572] |
| Normal analysis [0 Å$^{-1}$ ≤ sin $\theta$ /$\lambda$ ≤ 1.42 Å$^{-1}$] ($N_{Parameters}$ = 0) | |
| $R_1$ [# of reflections] | 0.0505 [69116] |
| $R_1$ ($I$ > 3$\sigma$) [# of reflections] | 0.0274 [47938] |
| GOF [# of reflections] | 1.06 [69116] |

We check for the presence of the inversion symmetry in $\alpha$-(BETS)$_2$I$_3$ (for the method, we referred to [S1]). Figure S2 shows the difference in intensities of the Friedel pairs [$I(hkl)$ and $I(\bar{h}\bar{k}\bar{l})$]. In $\alpha$-(ET)$_2$I$_3$ at 30 K, there is no inversion center. In this case, when comparing the intensities of the Friedel pairs for $|F_o|^2 (\propto I)$ and $|F_c|^2$ [Fig. S2(b)], $|F_o(hkl)|^2 - |F_o(\bar{h}\bar{k}\bar{l})|^2$ and $|F_c(hkl)|^2 - |F_c(\bar{h}\bar{k}\bar{l})|^2$ has a positive correlation, whose slope (red line) is ~1. The red line indicates the result of fitting by a linear function for each point so as to pass through the origin. This result indicates the lack of the inversion center, i.e., the space group is $P1$. When the result of $|F_o(hkl)|^2 - |F_o(\bar{h}\bar{k}\bar{l})|^2$ at 150 K is plotted using the $|F_c(hkl)|^2 - |F_c(\bar{h}\bar{k}\bar{l})|^2$ at 30 K on the horizontal axis [Fig. S2(a)], the slope is almost zero, which indicates the existence of the inversion center, i.e., the space group is $P\bar{1}$. We applied this analogy to $\alpha$-(BETS)$_2$I$_3$. When the crystal structural analysis of $\alpha$-(BETS)$_2$I$_3$ at 30 K was performed assuming $P1$ [Fig. S2(d)], we found that the slope was almost zero. This result indicates that the inversion center exists in the low-temperature phase, which is consistent with the result of the $^{13}$C NMR experiment [Fig. 1(c)].

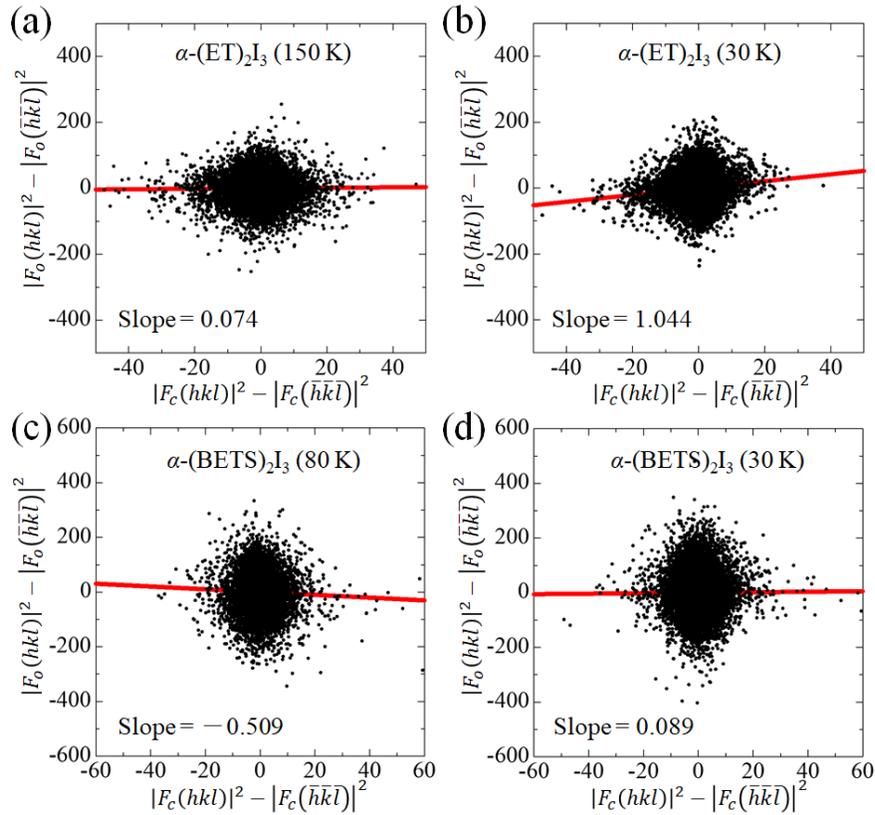

FIG. S2. Difference in intensities of the Friedel pairs. $|F_c|^2$ in (a) and (b) is the result of the crystal structural analysis of $\alpha$-(ET)$_2$I$_3$ at 30 K, in which $P1$ (no inversion) is assumed for the space group. $|F_c|^2$ in (c) and (d) is the result of the crystal structural analysis of $\alpha$-(BETS)$_2$I$_3$ at 30 K, in which $P1$ (no inversion) is assumed for the space group. In these figures, only reflections which have the intensity of less than 0.25% of the maximum intensity are plotted. The red line indicates the result of fitting by a linear function for each point so as to pass through the origin.

Table S4 shows the comparison of lattice parameters between this work and previous reports [S2,S3]. Figure S3 shows the comparison of intra-molecular bond length in α-(BETS)$_2$I$_3$ between our results at 80 K and 0 GPa and the result of Ref. [S3] at room temperature and 0.7 GPa. Only the length at the central C=C bond of molecule A does not match within each error range. As a result, the charge of molecule A deviates greatly from the tendency in our results [Fig. S3(d)].

Table S4. Comparison of lattice parameters with previous reports.

| Pressure [GPa] | Temp. [K] | $a$ [Å] | $b$ [Å] | $c$ [Å] | $\alpha$ [deg.] | $\beta$ [deg.] | $\gamma$ [deg.] | Volume [Å$^3$] | Ref. |
|---|---|---|---|---|---|---|---|---|---|
| 0 | R.T. | 9.209 | 10.816 | 17.777 | 96.63 | 97.89 | 90.69 | 1741.5 | [S2] |
| 0 | R.T. | 9.230 | 10.842 | 17.785 | 96.700 | 97.797 | 90.646 | 1750.6 | [S3] |
| 0.7 | R.T. | 9.069 | 10.727 | 17.675 | 96.508 | 97.625 | 90.631 | 1692.79 | [S3] |
| 0 | 80 | 9.0996(2) | 10.7301(2) | 17.7313(12) | 96.356(7) | 97.739(7) | 90.777(6) | 1704.25(13) | This work |
| 0 | 30 | 9.0922(2) | 10.7221(2) | 17.7377(12) | 96.310(7) | 97.706(7) | 90.794(6) | 1702.49(13) | This work |

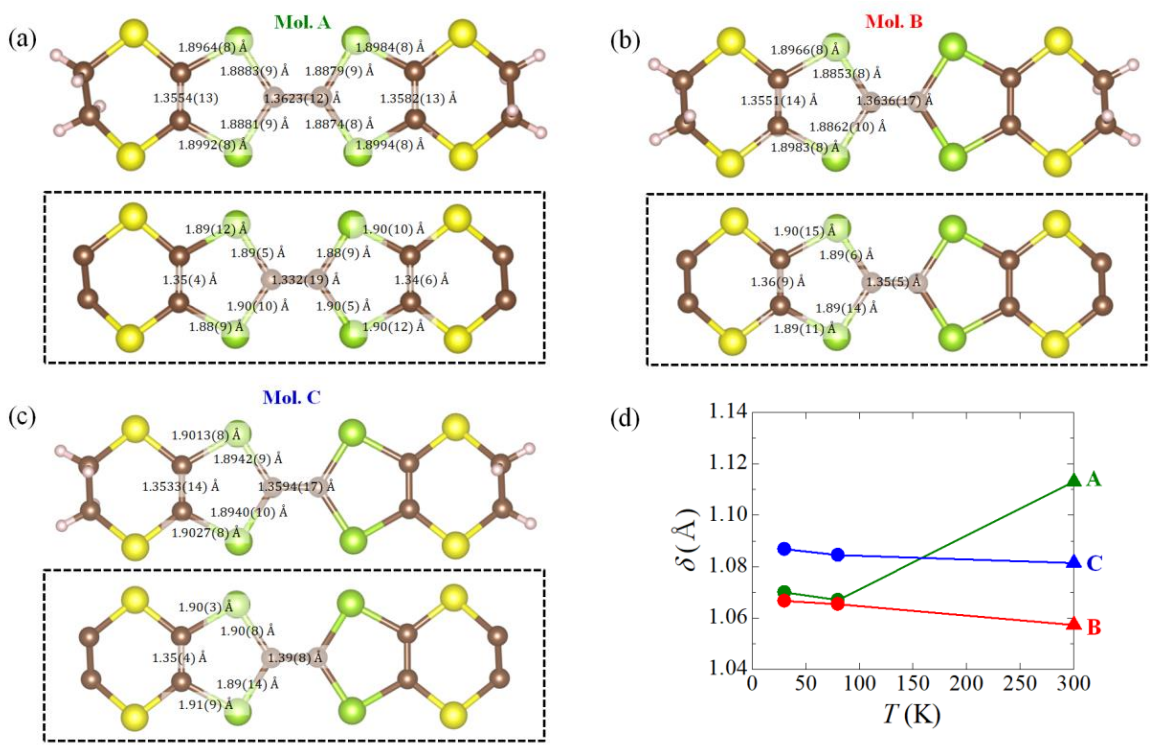

FIG. S3. Comparison of intramolecular bond length in (a) molecule A, (b) molecule B, and (c) molecule C, respectively. The upper molecules show our result at 80 K and 0 GPa. The lower molecules (inside the dotted rectangle) show the result of Ref. [S3] at room temperature and 0.7 GPa. (d) Temperature dependence of the $\delta$ value of BETS in α-(BETS)$_2$I$_3$, which is shown in Fig. 2(d) in the main manuscript. Circle and triangle indicate our result and the result of Ref. [S3], respectively.

Figures S4 and S5 show the results of the crystal structural analysis of α-(BETS)$_2$I$_3$ at 80 K and 30 K. The inter-molecular distances in the *a*-axis direction hardly changed between 80 K and 30 K (the ratio of change is 0.16% or less) (Fig. S4). The angles between the BETS molecules also hardly changed between 80 K and 30 K (Fig. S5). As a reference, the results of α-(ET)$_2$I$_3$ at 150 K and 30 K are shown in a dotted box in Fig. S5. In addition, we investigated the I–H distance between donor molecules and terminal iodine in α-(ET)$_2$I$_3$ and α-(BETS)$_2$I$_3$ because the importance of the hydrogen bonding between donor molecules and terminal iodine in I$_3$ molecules at the charge ordering phase transition of α-(ET)$_2$I$_3$ has been pointed out by P. Alemany *et al.* [S4]. Figures S6(a) and S6(b) show the temperature dependence of the I–H distances of α-(ET)$_2$I$_3$ and α-(BETS)$_2$I$_3$, respectively. The labeling of the I–H distance is shown in Fig. S4. Clear changes of the I–H distance are confirmed at the metal-insulator transition in α-(ET)$_2$I$_3$, which is consistent with the previous report [S4]. On the other hand, no changes are confirmed at the metal-insulator crossover in α-(BETS)$_2$I$_3$, which indicates the absence of the charge ordering in the LT phase of α-(BETS)$_2$I$_3$.

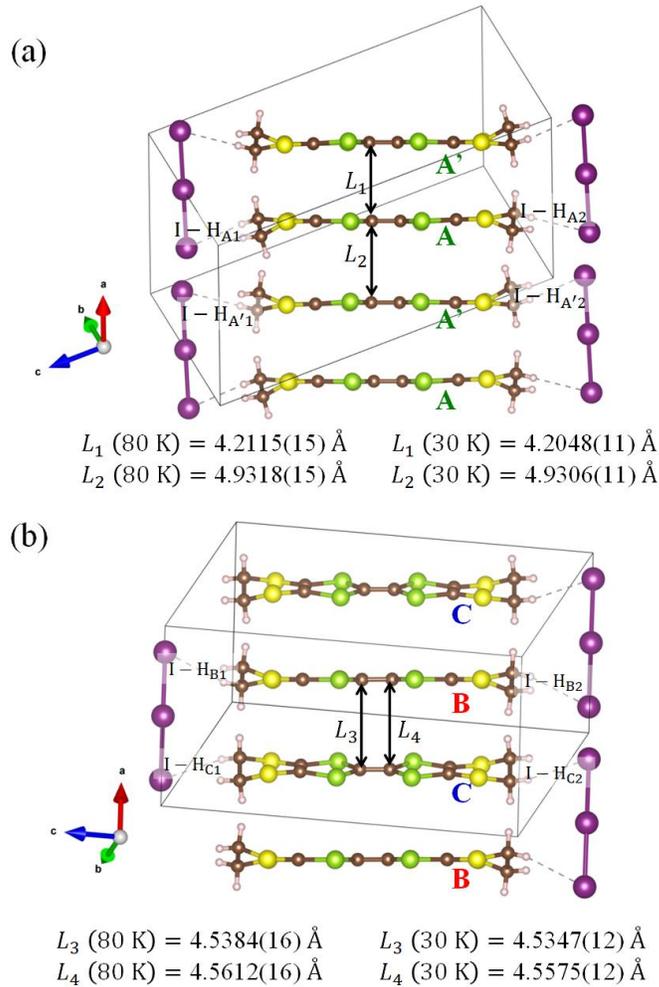

(a)

$L_1$ (80 K) = 4.2115(15) Å    $L_1$ (30 K) = 4.2048(11) Å
$L_2$ (80 K) = 4.9318(15) Å    $L_2$ (30 K) = 4.9306(11) Å

(b)

$L_3$ (80 K) = 4.5384(16) Å    $L_3$ (30 K) = 4.5347(12) Å
$L_4$ (80 K) = 4.5612(16) Å    $L_4$ (30 K) = 4.5575(12) Å

FIG. S4. Distances between the BETS molecules in α-(BETS)$_2$I$_3$ at 80 K and 30 K.

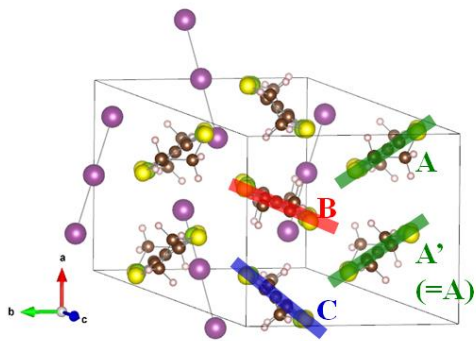

$$\cos\theta_{BC} = \boldsymbol{n}_B \cdot \boldsymbol{n}_C \qquad \theta_{BC}(80\text{ K}) = 12.545°$$
$$(|\boldsymbol{n}_B| = |\boldsymbol{n}_C| = 1) \qquad \theta_{BC}(30\text{ K}) = 12.521°$$

*cf.*) $\alpha$-(ET)$_2$I$_3$

$\theta_{AA'}(150\text{ K}) = 0°$ $\qquad \theta_{BC}(150\text{ K}) = 12.387°$
$\theta_{AA'}(30\text{ K}) = 1.969°$ $\qquad \theta_{BC}(30\text{ K}) = 13.179°$

FIG. S5. Angles between the BETS molecules in $\alpha$-(BETS)$_2$I$_3$ at 80 K and 30 K. The plane through which central ten atoms (six C and four Se) in BETS pass was calculated by the method of least squares. The angle between two planes was calculated from the normal vectors $\boldsymbol{n}$ of the planes.

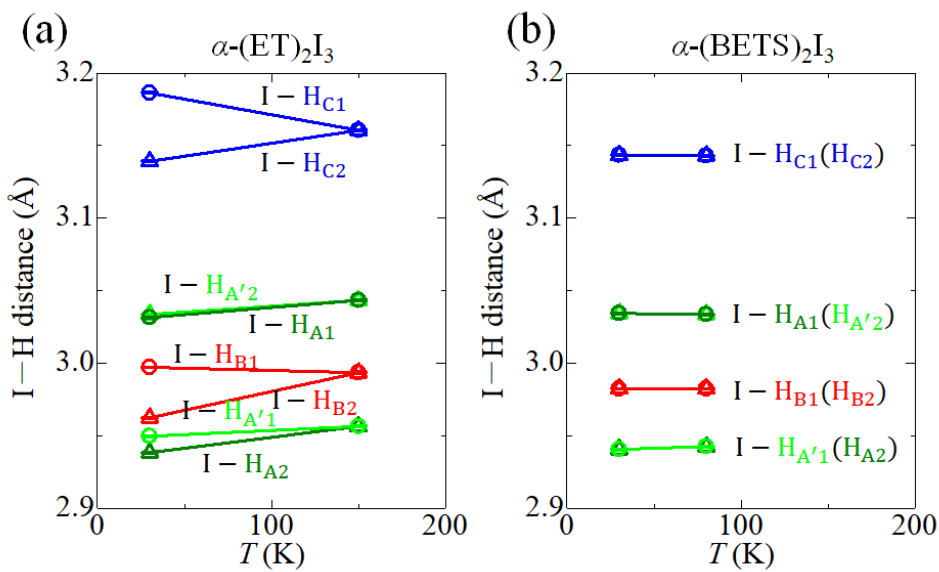

FIG. S6. Temperature dependence of the I–H distances between donor molecules and terminal iodine in (a) $\alpha$-(ET)$_2$I$_3$ and (b) $\alpha$-(BETS)$_2$I$_3$. The labeling of the I–H distance is shown in Fig. S4.

## 2. Electron density analysis

To investigate the valence electron density (VED) distribution in $\alpha$-(ET)$_2$I$_3$ and $\alpha$-(BETS)$_2$I$_3$, we performed the electron density analysis using a core differential Fourier synthesis (CDFS) method [S5]. The equation of the inverse Fourier transform by the CDFS method is described as

$$\rho_v(\boldsymbol{r}) = \frac{1}{V}\sum_{\boldsymbol{K}}\left[\left(|F_o(\boldsymbol{K})|P - \left|\sum_j f_j^{\text{core}} T_j e^{-i\boldsymbol{K}\cdot\boldsymbol{r}_j}\right|P^{\text{core}}\right)e^{i\boldsymbol{K}\cdot\boldsymbol{r}}\right] + \frac{n_v}{V}.$$

(S1)

Here, $\rho_v(\boldsymbol{r})$ corresponds the VED, $V$ is the cell volume, $\boldsymbol{r}_j$ is the $j^{\text{th}}$ atomic position, $T_j$ is the $j^{\text{th}}$ atomic displacement parameter. $f_j^{\text{core}}$ is the $j^{\text{th}}$ atomic scattering factor with only the core electrons contribution [S6,S7], which corresponds to the blue line in Fig. S7. $P$ and $P^{\text{core}}$ is the phase term, which is calculated as $P = F_c(\boldsymbol{K})/|F_c(\boldsymbol{K})|$ and $P^{\text{core}} = F_c^{\text{core}}(\boldsymbol{K})/|F_c^{\text{core}}(\boldsymbol{K})|$, respectively. $n_v$ is the total number of valence electrons contained in the unit cell.

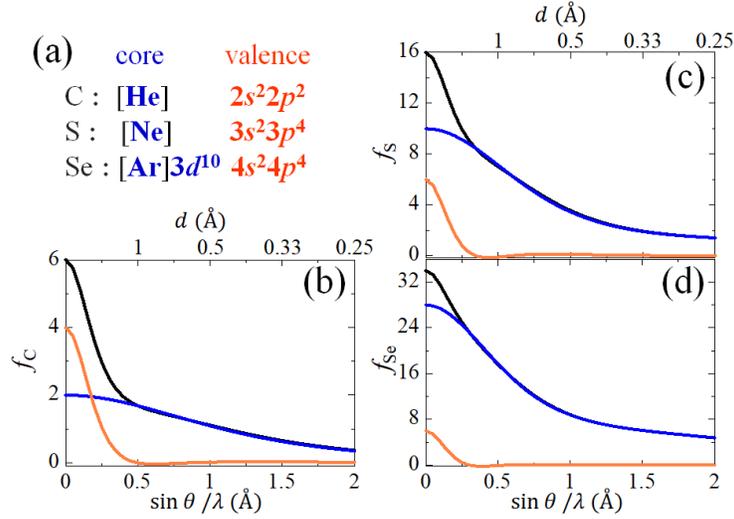

FIG. S7. (a) Electron configurations of C, S, and Se. Atomic scattering factor of (b) C, (c) S, and (d) Se [S7]. Black, blue, and orange lines indicate the contribution of the total, core, and valence electrons, respectively.

Figure S8 show the VED distributions of molecule B and C. Figure S9 shows the difference between the VED distributions of molecule A (hole-poor) and A' (hole-rich) in $\alpha$-(ET)$_2$I$_3$ at 30 K, where there are no clear differences in the electron density distribution. Figure S10 shows the VED distributions calculated by the diffraction data with the different $(\sin\theta/\lambda)_{\text{max}}$ range corresponding to the real space resolution $d(=\lambda/2\sin\theta)$. These VED distributions are qualitatively the same regardless of the resolution, but especially in Fig. S10(c), the electron density appears to be disturbed due to the influence of high-angle reflections, whose intensities are weak (i.e. bad *S/N* ratio).

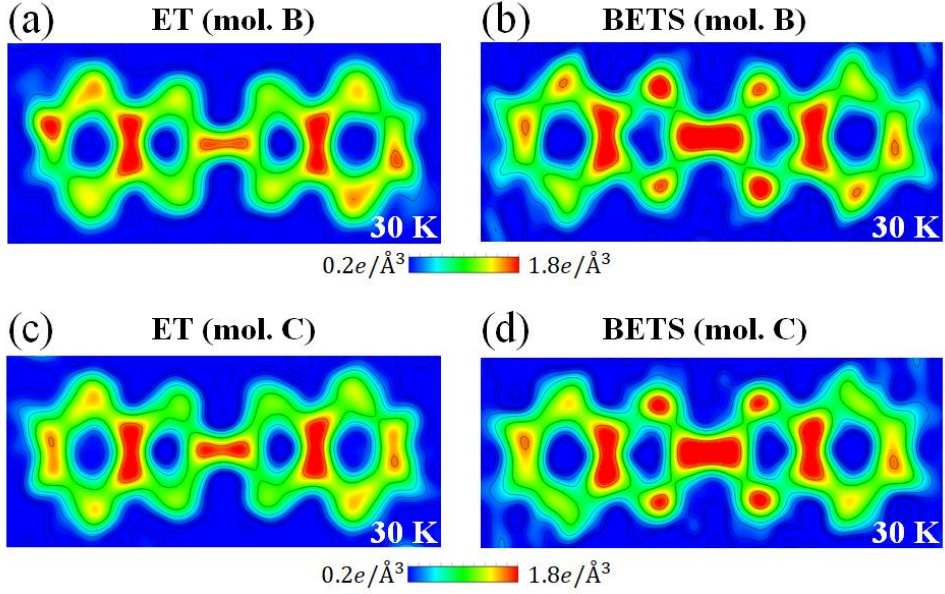

FIG. S8. VED distribution of (a) molecule B in $\alpha$-(ET)$_2$I$_3$ at 30 K, (b) molecule B in $\alpha$-(BETS)$_2$I$_3$ at 30 K, (c) molecule C in $\alpha$-(ET)$_2$I$_3$ at 30 K, and (d) molecule C in $\alpha$-(BETS)$_2$I$_3$ at 30 K, which are obtained from the CDFS analysis. These VED distributions are calculated by the diffraction data in the limit $0\ \text{Å}^{-1} \leq \sin\theta/\lambda \leq 0.5\ \text{Å}^{-1}$.

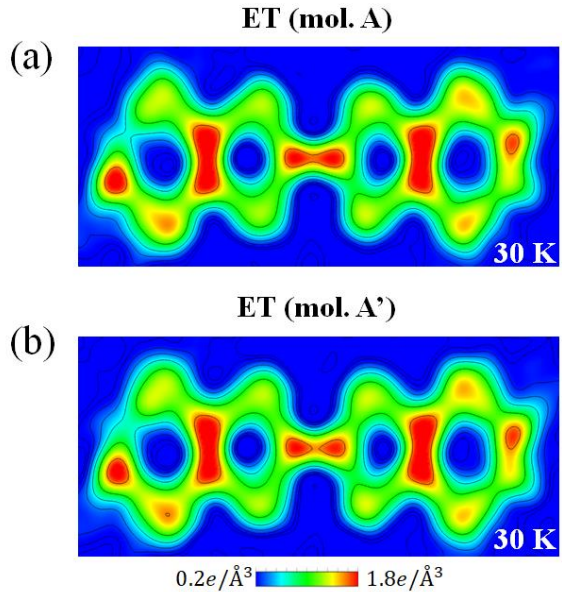

FIG. S9. Difference of the experimental VED distributions between (a) molecule A (hole-poor) and (b) molecule A' (hole-rich) in $\alpha$-(ET)$_2$I$_3$ at 30 K. These VED distributions are calculated by the diffraction data in the limit $0\ \text{Å}^{-1} \leq \sin\theta/\lambda \leq 0.5\ \text{Å}^{-1}$.

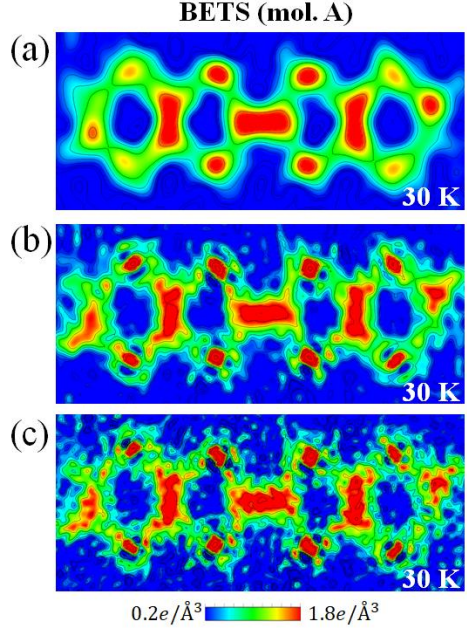

FIG. S10. VED distribution obtained by the CDFS analysis in $\alpha$-(BETS)$_2$I$_3$ at 30 K, which are calculated by the diffraction data in the limit (a) $0 \text{ Å}^{-1} \leq \sin\theta/\lambda \leq 0.5 \text{ Å}^{-1}$, (b) $0 \text{ Å}^{-1} \leq \sin\theta/\lambda \leq 1 \text{ Å}^{-1}$, and (c) $0 \text{ Å}^{-1} \leq \sin\theta/\lambda \leq 1.42 \text{ Å}^{-1}$, respectively.

### 3. Computational details for electronic structures

The band structures shown in Fig. 4 are calculated by first-principles density functional theory within the generalized gradient approximation (GGA) to the exchange-correlation functional proposed by PBE [S8]. Kohn-Sham equations are self-consistently solved in a scalar-relativistic fashion using the all-electron full-potential linearized augmented plane wave (FLAPW) method implemented in the QMD-FLAPW12 code [S9-S11]. The LAPW basis functions in the interstitial region have a cutoff energy of 20.3 Ry. The angular momentum expansion inside the muffin-tin (MT) sphere is truncated at $l = 8$ for all the atoms. The cutoff energy for the potential and density is 282 Ry. The MT sphere radii are set as 1.26, 0.75, 2.00, and 2.27 Bohr for C, H, S, and Se atoms, respectively. The electronic states up to C $(2s)^2$, S $(2p)^6$, Se $(3p)^6$, and I $(4d)^{10}$ are treated as core electrons, which are predominantly confined to the MT spheres. $\boldsymbol{k}$-point meshes used are $6 \times 6 \times 2$ for both self-consistent field and the density of states (DOS) calculations. The local density of states (LDOS) shown in Fig.4(b) are obtained as a summation of projected DOS on C $p$, S $p$, and Se $p$ states in the respective monomer units. We used a high-dense $\boldsymbol{k}$-mesh for plotting the 3D band structures shown in Fig.4(c).

The band structure calculations are also performed using the pseudopotential method based on the projector augmented wave (PAW) formalism [S12] with plane wave basis sets implemented in the Quantum Espresso (version 6.3) [S13]. The results of the scalar relativistic calculations are fairly in agreement with each other. The cutoff energies for plane waves and charge densities are set to be 55 (48) and 488 (488) Ry in the scalar (full) relativistic calculations, respectively. We used $4 \times 4 \times 2$ uniform $\boldsymbol{k}$-point mesh with a gaussian smearing method during self-consistent loops. In both scalar and full relativistic pseudopotentials, the valence configurations of the pseudopotentials are C: $(2s)^2$ $(2p)^2$, H: $(1s)^1$, S: $(3s)^2$ $(3p)^4$, Se: $(4s)^2$ $(4p)^4$ $(3d)^{10}$, and I: $(5s)^2$ $(5p)^5$ $(4d)^{10}$. The pseudopotentials are generated using "atomic" code by A. Dal Corso v.6.3 [S14], where the pseudization algorithm proposed by Troullier and Martins [S15] and non-linear core correction [S16] are used.

## 4. Computational details for $Z_2$ topological invariants

The density functional calculations of $Z_2$ topological invariants are performed by computing parity eigenvalues and Fukui-Hatsugai method using OpenMX code [S17]. The computational details are summarized in Table S6. We used GGA-PBE as the exchange correlation functional [S8]. We adapt norm-conserving pseudopotentials with an energy cutoff of 300 Ry for the charge density, including the $2s$ and $2p$-states as states for C; $1s$ for H; $5s$ and $5p$ for I; $3s$ and $3p$ for S; $4s$ and $4p$ for Se. The wavefunctions were expanded by the linear combination of numerical pseudoatomic orbitals [S18,S19]. Spin-orbit interactions were included by a fully relativistic $j$-dependent pseudopotential, where $j$ is the total angular momentum. The numerical pseudo atomic orbitals are as follows: the numbers of the $s$-, $p$-, and $d$-character orbitals are 2, 2, and 1, respectively, for C, S, and Se; 2, 1, and 0, respectively, for H; 2, 2, and 2, respectively, for I. The cutoff radii of C, S, Se, H, and I are 5.0, 7.0, 7.0, 5.0, and 5.0, respectively, in units of Bohr. The regular $k$-point mesh $7 \times 7 \times 7$ was used for self-consistent field calculations. The $20 \times 20$ $k$-point mesh for four independent two-dimensional tori was used for computing $Z_2$ topological invariants by Fukui-Hatsugai method [S20].

Table S6. The computational details for the $Z_2$ topological invariants

| Pseudo atomic basis set and cutoff radius (The number after $s$, $p$, $d$ is the radial function multiplicity of each angular momentum component.) | Atom | Basis set | Cutoff radius (Bohr) |
|---|---|---|---|
| | C | $s^2p^2d^1$ | 5.0 |
| | H | $s^2p^1$ | 5.0 |
| | I | $s^2p^2d^2$ | 5.0 |
| | S | $s^2p^2d^1$ | 7.0 |
| | Se | $s^2p^2d^1$ | 7.0 |
| $k$-space sampling points for self-consistent field calculations | $7 \times 7 \times 7$ | | |
| Cutoff energy | 300 Ry | | |
| $k$-space sampling points for Fukui-Hatsugai method | $20 \times 20 \times 1$ | | |

Table S7. $Z_2$ topological invariants of $\alpha$-(BETS)$_2$I$_3$ and $\alpha$-(ET)$_2$I$_3$

| Molecule | Temperature | Space group | $Z_2$ topological invariants |
|---|---|---|---|
| $\alpha$-(BETS)$_2$I$_3$ | 30 K | $P\bar{1}$ | (0; 0 0 1) |
| | 80 K | $P\bar{1}$ | (0; 0 0 1) |
| $\alpha$-(ET)$_2$I$_3$ | 30 K | $P1$ | (0; 0 0 0) |
| | 150 K | $P\bar{1}$ | (0; 0 0 1) |